\documentstyle[preprint,aps]{revtex}
\begin{document}
\draft
\newcommand{\be}{\begin{equation}}
\newcommand{\ee}{\end{equation}}
\begin{title}
{\bf SOLITON STABILITY IN SYSTEMS OF TWO\\
           REAL SCALAR FIELDS}
\end{title}

\author{D. Bazeia, J. R. S. Nascimento, R. F. Ribeiro, and D. Toledo}

\address{Departamento de F\'\i sica, Universidade Federal da Para\'\i ba\\
Caixa Postal 5008, 58051-970 Jo\~ao Pessoa, Para\'\i ba, Brazil}

\maketitle

\begin{abstract}

In this paper we consider a class of systems of two coupled real scalar fields
in bidimensional spacetime, with the main motivation of studying classical
or linear stability of soliton solutions. Firstly, we present the class of
systems and comment on the topological profile of soliton solutions one can
find from the first-order equations that solve the equations of motion. After
doing that, we follow the standard approach to classical stability to
introduce the main steps one needs to obtain the spectra of Schr\"odinger
operators that appear in this class of systems. We consider a specific system,
from which we illustrate the general calculations and present some
analytical results. We also consider another system, more general, and
we present another investigation, that introduces new results and offers a
comparison with the former investigations.
\end{abstract}
\vskip 1cm

\pacs{PACS numbers: }


\section{Introduction}

Lagrangian systems described by coupled scalar fields are gaining renewed
attention recently. In the case of real fields, in particular, the work
presented in \cite{bds95,bsa96,brs96a,brs96b} has introduced a specific class
of systems of two coupled real scalar fields. In distinction to older models,
the class of systems introduced in the above papers is very peculiar, at least
in bidimensional spacetime, where it presents the following general properties:
firstly, the corresponding equations of motion are solved by field
configurations obeying first-order differential equations; secondly, the
classical configurations that solve the first-order equations present minimum
energy and are classically or linearly stable. On the other hand, the
first-order differential equations can be seen as a dynamical system, and so
we can take advantage of all the mathematical tools available to dynamical
systems to deal with the set of first-order equations and, consequently,
with solutions to the equations of motion.

For the specific issue concerning classical or linear stability, in
\cite{bsa96,brs96b} a general way of investigating stability was presented.
In this case, the investigation relies essentially on proving that the
associate Schr\"odinger operator is positive semi-definite. This procedure
of implementing stability investigations \cite{jac77} is in distinction to
the so-called standard approach \cite{jac77,raj82}, where one usually obtains
the complete set of eigenvalues of the Schr\"odinger operator.

However, we know that the standard approach is important since it sets forward
results one needs to implement quantum corrections, since in this case one
has to know explicitly the spectrum of the corresponding Schr\"odinger
operator \cite{jac77,raj82}. We then think that a standard investigation of
classical stability of systems of two real scalar fields belonging to the
class of systems already introduced in \cite{bds95,bsa96,brs96a,brs96b} is
welcome, not only to present comparison with former investigations but also
to unveil the full spectra of the corresponding Schr\"odinger operators.
Within this context, the main motivation of the present paper is to deal
with issues concerning finding spectra of Schr\"odinger operators that
naturally appear in the standard approach to linear stability of systems
belonging to the class of coupled fields introduced in
\cite{bds95,bsa96,brs96a,brs96b}. As we are going to show, in this class of
systems the Schr\"odinger operators can be written in terms of first-order
operators, and this property eases the calculation toward obtaining the
corresponding energy spectra. 

Evidently, to implement the standard approach to linear stability we have to
deal with specific systems. However, to make the present investigation as
general as possible, we have organized the present work as follows. In the
next Sec.~{\ref{sec:systems}} we present the class of systems of two coupled
real scalar fields and comment on the topological profile of the classical
configurations. In Sec.~{\ref{sec:stability}} we investigate linear stability
within the standard approach, and we shed new light on the issue concerning
unveiling the full spectra of Schr\"odinger operators that appear in the
class of systems introduced in the former section. In order to illustrate
the general calculations, in Sec.~{\ref{sec:example}} we examine a particular
system of two coupled fields. We end this paper in Sec.~{\ref{sec:general}},
where we present another investigation, in which we deal with a more general
system, not belonging to the class of systems we shall introduce in
Sec.~{\ref{sec:systems}}. Each section contains some comments and conclusions.

\section{Systems of Two Real Scalar Fields}
\label{sec:systems}

Let us start with the Lagrangian density
\be
{\cal L}= \frac{1}{2} \partial_{\alpha}\phi\partial^{\alpha}\phi +
\frac{1}{2} \partial_{\alpha}\chi\partial^{\alpha}\chi- 
U,
\ee
where $U=U(\phi,\chi)$ is the potential, in general a nonlinear function of
the two fields $\phi$ and $\chi$. Here we are using natural units, and so
$\hbar=c=1$, and the metric is such that $x^{\alpha}=(t,x)$ and
$x_{\alpha}=(t,-x)$. The class of systems introduced in
\cite{bds95,bsa96,brs96a,brs96b} is defined by the following potential
\be
\label{eq:pot}
U(\phi,\chi)=\frac{1}{2}H^2_{\phi}+\frac{1}{2}H^2_{\chi},
\ee
where $H=H(\phi,\chi)$ is a smooth but otherwise arbitrary
function of the fields $\phi$ and $\chi$, and $H_{\phi}=\partial
H/\partial\phi$, $H_{\chi}=\partial H/\partial\chi$.

The Euler-Lagrange equations, that is, the equations of motion that follow
from the above system are given by
\be
\label{eq:soeqnst1}
\frac{\partial^2\phi}{\partial t^2}-\frac{\partial^2\phi}{\partial x^2}+
H_{\phi}H_{\phi\phi}+H_{\chi}H_{\phi\chi}=0,
\ee
and 
\be
\label{eq:soeqnst2}
\frac{\partial^2\chi}{\partial t^2}-\frac{\partial^2\chi}{\partial x^2}+
H_{\phi}H_{\chi\phi}+H_{\chi}H_{\chi\chi}=0,
\ee
and for static field configurations they change to
\be
\label{eq:soeqnsx1}
\frac{d^2\phi}{dx^2}=H_{\phi}H_{\phi\phi}+
H_{\chi}H_{\phi\chi},
\ee
and
\be
\label{eq:soeqnsx2}
\frac{d^2\chi}{dx^2}=H_{\phi}H_{\chi\phi}+
H_{\chi}H_{\chi\chi}.
\ee
Recall that static field configurations are configurations written in their
rest reference frame.

For static field configurations, the energy can be cast to the form
$E=E_M+E'$, where
\be
E_M=H(\phi(\infty),\chi(\infty))-
H(\phi(-\infty),\chi(-\infty)),
\ee
and
\be
E'=\frac{1}{2}\int^{\infty}_{-\infty}
\Biggl[\left(\frac{d\phi}{dx}-H_{\phi}\right)^2+
\left(\frac{d\chi}{dx}-H_{\chi}\right)^2\Biggr].
\ee
As we have already learned from \cite{brs96b}, we impose the conditions 
\be
\label{eq:foeqns}
\frac{d\phi}{dx}=H_{\phi},\hspace{1cm}
\frac{d\chi}{dx}=H_{\chi},
\ee
and in this case we see that the energy gets to its lower bound $E_M$, and
the above first-order equations $(\ref{eq:foeqns})$ solve the corresponding
equations of motion $(\ref{eq:soeqnsx1})$ and $(\ref{eq:soeqnsx2})$.

Before investigating classical stability, let us first comment on the issue
concerning topological properties of soliton solutions. From the above
calculations on energy of the corresponding static fields, we see that field
configurations obeying the pair of first-order equations present minimum
energy that only depends on the difference between the two asymptotic
behaviors of $H(\phi,\chi)$. In this case, classical pairs of static field
configurations having finite but nonzero energy must necessarily obey the
topological property of connecting distinct minima of the corresponding
potential. As one knows, this type of field configurations are named
topological solutions \cite{raj82}. On the other hand, one also knows that
systems of two coupled scalar fields may present nontopological solutions
\cite{raj82}, and in this case the field configurations must have the same
asymptotic behavior. In the above class of systems, however, static field
configurations obeying the first-order differential equations
$(\ref{eq:foeqns})$ can not have nontopological profile, because this would
give zero energy to the pair of solutions, and zero is the energy value of
the vacua states. This result is very interesting, because it leads us to the
fact that the search for classical solutions of the equations of motion via
the first-order equations $(\ref{eq:foeqns})$ evidently does not give us the
full set of physical solutions, and certainly no nontopological solution.

Another interesting point relies on considering the set of first-order
equations as a dynamical system. This procedure allows unveiling the full
set of singular points, together with stability properties of each one of
these points. As we can see from the potential $(\ref{eq:pot})$, the set of
singular points is identified with the vacuum manifold of the corresponding
field theory. Then, once we know the vacuum manifold and the classification
of each one of its points as stable, unstable and saddle points, we have
everything one needs to deal with finding explicit soliton solutions. The
route toward finding explicit solutions can,
for instance, follow the trial orbit method introduced in
\cite{raj79}. However, since here we are dealing with first-order equations
this trial orbit method becomes easer
to be implemented than it is in the original work \cite{raj79}. 

\section{Classical Stability}
\label{sec:stability}

Let us now focus attention on the issue concerning classical or linear
stability. In this case we consider $\bar{\phi}=\bar{\phi}(x)$ and
$\bar{\chi}=\bar{\chi}(x)$ as a pair of static solutions to the above
first-order differential equations.  Here we consider
$\phi(x,t)=\bar{\phi}(x)+ \eta_n(x)\cos(w_n t)$ and 
$\chi(x,t)=\bar{\chi}(x)+ \xi_n(x)\cos(w_nt)$ in order to get, from the
equations of motion $(\ref{eq:soeqnst1})$ and $(\ref{eq:soeqnst2})$,
working up to first order in the fluctuations,
\begin{eqnarray}
\label{eq:schro}
S_2\pmatrix{\eta_n\cr \xi_n}=
w^2_n\pmatrix{\eta_n\cr \xi_n}.
\end{eqnarray}
This is a Schr\"odinger equation, and $S_2$ is the Schr\"odinger operator,
given by
\be
S_2=-\frac{d^2}{dx^2}+ V,
\ee
where the potential $V$ has the form
\begin{eqnarray}
V=\pmatrix{V_{11}& V_{12}\cr
V_{21}& V_{22} },
\end{eqnarray}
and the matrix elements can be written as
\be
V_{11}=\bar{H}^2_{\phi\phi}+\bar{H}^2_{\chi\phi}+
       \bar{H}_{\phi}\bar{H}_{\phi\phi\phi}+
       \bar{H}_{\chi}\bar{H}_{\phi\phi\chi},
\ee
and
\be
V_{12}=V_{21}=\bar{H}_{\phi\phi}\bar{H}_{\phi\chi}+
              \bar{H}_{\chi\chi}\bar{H}_{\phi\chi}+
              \bar{H}_{\phi}\bar{H}_{\phi\phi\chi}+
              \bar{H}_{\chi}\bar{H}_{\phi\chi\chi},
\ee
and
\be
V_{22}=\bar{H}^2_{\chi\chi}+\bar{H}^2_{\phi\chi}+
       \bar{H}_{\phi}\bar{H}_{\phi\chi\chi}+
       \bar{H}_{\chi}\bar{H}_{\chi\chi\chi}.
\ee
In the above expressions a bar over $H$ means that the corresponding quantity
has to be calculated at the classical static values $\phi=\bar{\phi}(x)$ and
$\chi=\bar{\chi}(x)$, and so $V_{ij}=V_{ij}(x)$, for $i,j=1,2$. 

On the other hand, as it was already shown in \cite{bsa96,brs96b}, from the
first-order equations we can introduce the first-order operators
\be
S^{\pm}_1=\pm\frac{d}{dx}+v,
\ee
where $v$ is given by
\begin{eqnarray}
v=\pmatrix{H_{\phi\phi} & H_{\phi\chi} \cr
H_{\phi\chi} & H_{\chi\chi} }.
\end{eqnarray}
Here it is not hard to check that these first-order operators are adjoint of
each other, and that $S_2=S^{+}_1S^{-}_1$. This is important because it proves
that the Schr\"odinger operator $S_2$ is positive semi-definite, and this is
the result one needs to ascertain that the pair of solutions $\bar{\phi}$ and
$\bar{\chi}$ is classically or linearly stable.

To investigate the spectrum of the Schr\"odinger operator $S_2$ we have to go
further into this problem, and here the main difficulty concerns solving the
Schr\"odinger equation $(\ref{eq:schro})$. The task is not immediate since in
the case of two coupled fields the fluctuations $\eta$ and $\xi$ are also
coupled, in general, and so the first issue we have to deal with concerns
diagonalizing the second-order Schr\"odinger operator, a calculation to be
done by finding the normal mode fluctuations.

In the above class of systems of two coupled fields, however, we take
advantage of the presence of the corresponding first-order operators
$S^{\pm}_1$, and so the task of finding the normal mode fluctuations is
greatly simplified. This is so because here we can deal with the simpler
task of just diagonalizing the $v$ matrix which appears in the first-order
operators. In this case the result allows writing
\begin{eqnarray}
\bar{S}^{\pm}_1=\pm\frac{d}{dx}+
\pmatrix{v_{+} & 0 \cr
0 & v_{-} },
\end{eqnarray}
where the diagonal elements are given by 
\be
v_{\pm}=\frac{1}{2}(\bar{H}_{\phi\phi}+\bar{H}_{\chi\chi})
\pm\Bigl[(1/4)(\bar{H}_{\phi\phi}-
\bar{H}_{\chi\chi})^2+\bar{H}^2_{\phi\chi}\Bigr]^{1/2}.
\ee

If we use the notation $\eta_{\pm}$ for the pair of normal mode fluctuations,
then we have to deal with the following Schr\"odinger equations
\be
\bar{S}_2^{\pm}\eta_{\pm}=w^2\eta_{\pm},
\ee
where the Schr\"odinger operators $\bar{S}_2^{\pm}$ are now given by
\be
\bar{S}^{\pm}_2 =-\frac{d^2}{dx^2}+V_{\pm},
\ee
with the potentials
\be
V_{\pm}=v^2_{\pm}+\frac{dv_{\pm}}{dx}.
\ee

Before paying attention to specific systems, let us reason a little more on
the issue concerning linear stability. As we can see from the above
investigation, to unveil the spectra of $\bar{S}^{\pm}_2$ we recognize that
the square root that appears in $v_{\pm}$ complicates the calculation, and
will certainly require numerical investigations. To circumvent this difficulty
and perhaps give explicit analytical results, we should focus attention on
avoiding the square root in $v_{\pm}$. Here we see that the simplest case
where the square root in $v_{\pm}$ desapears is when $\bar{H}_{\phi\chi}=0$.
However, since we are dealing with systems of two coupled scalar fields, to
work with nontrivial systems we must have $H_{\phi\chi}$ nonzero, in order to
account for interactions between the two fields. In this way, to get to
$\bar{H}_{\phi\chi}=0$, the pair of classical solutions $\bar{\phi}$ and
$\bar{\chi}$ must be very specific. However, in general the quantity
$\bar{H}_{\phi\chi}$ does contribute, and so we must first deal with the
quantity
\be
\label{eq:radical}
R=\frac{1}{4}(H_{\phi\phi}-H_{\chi\chi})^2+
H^2_{\phi\chi},
\ee
which appears inside the square root in $v_{\pm}$. Here we should work out a
way of avoiding the classical pair $\bar{\phi}$ and $\bar{\chi}$ to remain
inside the square root. This reasoning will become clearer in the following,
where we deal with a specific system.

\section{An Example}
\label{sec:example}

As a particular example, let us consider the system defined by the function
\be
\label{eq:example}
H(\phi,\chi)=\lambda\left(\frac{1}{3}\phi^3-a^2\phi\right)+
\frac{1}{2}\mu\phi\chi^2.
\ee
In this case the potential that specifies the system has the form
\be
\label{eq:poten}
U(\phi,\chi)=\frac{1}{2}\lambda^2(\phi^2-a^2)^2+
\frac{1}{2}\lambda\mu(\phi^2-a^2)\chi^2+
\frac{1}{8}\mu^2\chi^4+\frac{1}{2}\mu^2\phi^2\chi^2.
\ee
Our main motivation to work with the above system is that
it is similar to the model already investigated in
\cite{raj79}, for which a standard stability investigation
was already done \cite{bca89}, and that it presents pairs of soliton
solutions \cite{bds95} that are very similar to the pairs considered in
\cite{bca89}. This motivation broadens with the fact that this specific
system was shown to be useful not only in field theory \cite{brs96a} but
also in condensed matter \cite{brs96b}.

In this case the first-order equations become
\be
\frac{d\phi}{dx}=\lambda (\phi^2-a^2)+\frac{1}{2}\mu\chi^2,
\ee
and
\be
\frac{d\chi}{dx}=\mu\phi\chi.
\ee
This system was already investiagted in \cite{bds95}, and some pairs of
soliton solutions were presented. In particular, a pair of solutions is
\be
\bar{\phi}_1(x)=-a\tanh(\lambda a x), \qquad \bar{\chi}_1(x)=0.
\ee
We recall that this pair of solutions introduces no rectrictions on the two
parameters $\lambda$ an $\mu$. Another pair of solutions is
\begin{eqnarray}
\bar{\phi}_2(x)&=&-a\tanh(\mu a x),\\
\bar{\chi}_2(x)&=&\pm a \sqrt{2\left( \frac{\lambda}{\mu}-1\right) }\,
{\rm{sech}}(\mu a x).
\end{eqnarray}
For this second pair of solutions the parameters $\lambda$ and $\mu$ are
restricted to satisfy $\lambda/\mu >1$. Note that the limit
$\lambda/\mu\to 1$ transforms the second pair of solutions into the first one.
Note also that for the second pair of solutions the field configurations obey
\be
\label{eq:orbit}
\bar{\phi}_2^2 +\frac{1}{2}\left(\frac{\lambda}{\mu}-
1\right)^{-1} \bar{\chi}_2^2= a^2.
\ee
We see that both pairs of solutions connect the points $(a,0)$ and $(-a,0)$
in the $(\phi,\chi)$ plane, the first by a straight line, and the second by
a elliptical line, as shown in Fig.1.

\begin{eqnarray}
\put (-30,0){\line(1,0){60}}
\qbezier (-30,0)(-20,15)(0,15)
\qbezier (30,0)(20,15)(0,15)\nonumber
\end{eqnarray}
\begin{center}
Fig.1. The two pairs of soliton solutions.
\end{center}

As we can see, these two pairs of soliton solutions belong to the same
topological sector, and present the same energy \cite{bds95,brs96b}.
Furthermore, the solutions are very similar to the pairs of classical
configurations investigated in \cite{bca89}, and so it seems
interesting to compare the present calculations with the ones there
introduced. Here we recall that we already know that the above
pairs of solutions are stable \cite{bsa96,brs96b},
while the pairs considered in \cite{bca89} were shown
to be unstable, at least in the region of parameters there considered. This is
a interesting result, since it shows that, in distinction to the class of
systems defined by the function $H=H(\phi,\chi)$, older systems like the one
presented in \cite{raj79} may perhaps have no classically stable soliton
solutions, and this is true at least in some region in parameter space.

To obtain the spectra of the corresponding Schr\"odinger operators we use
$H(\phi,\chi)$ given by $(\ref{eq:example})$ to write
$H_{\phi\phi}=2\lambda\phi$, $H_{\chi\chi}=\mu\phi$, and
$H_{\phi\chi}=\mu\chi$. For the first pair of solutions we get
$\bar{H}_{\phi\chi}=0$, and so the equations for the fluctuations are
already decoupled. This is very specific, and appears because of the
classical value $\bar{\chi}=0$. In this case we get
\be
v_{+}=-2\lambda a\tanh(\lambda ax),
\ee
and
\be
v_{-}=-\mu a\tanh(\lambda ax).
\ee

For the second pair of solutions the fluctuations are
coupled, and so firstly we consider $R$ as given
by Eq.$(\ref{eq:radical})$. Here we have
\be
R=\mu^2\left(\frac{\lambda}{\mu}-
\frac{1}{2}\right)^2\bar{\phi}_2^2+\mu^2\bar{\chi}_2^2.
\ee
We use the orbit $(\ref{eq:orbit})$ to rewrite the above
quantity as
\be
R=\mu^2\left(\frac{\lambda}{\mu}-
\frac{3}{2}\right)^2\bar{\phi}_2^2+
2\mu^2\left(\frac{\lambda}{\mu}-1\right)a^2.
\ee
In this case the parameters obey $\lambda/\mu >1$, and so the only way of
getting rid of the classical field from the square root is by setting to zero
the coefficient of $\bar{\phi}$. Here we get the relation $\lambda=(3/2)\mu$,
and this is a very interesting point in parameter space since it equals the
amplitude of the two classical configurations, as we can immediately see from
the second pair of solutions; also, the value $\lambda=(3/2)\mu$ changes the
in general elliptical profile of the orbit $(\ref{eq:orbit})$ to the very
particular case of a circular one.

For the first pair of solutions, the Schr\"odinger operators corresponding to
fluctuations about the $\phi$ and $\chi$ fields are given by, respectively,
\be
\bar{S}^{(1,1)}_2=-\frac{d^2}{dx^2}+4\lambda^2 a^2-
6\lambda^2a^2{\rm sech}^2(\lambda ax),
\ee
and
\be
\bar{S}^{(1,2)}_2=-\frac{d^2}{dx^2}+\mu^2a^2-
\mu(\lambda+\mu)a^2 {\rm sech}^2(\lambda ax).
\ee
This system was already investigated in \cite{bds95}. From the results there
obtained we see that, for $\lambda=(3/2)\mu$, fluctuations about the $\phi$
field presents the zero mode, and a bound state at the value
$w^2=(3/4)9\mu^2a^2$, with the continuum starting at $9\mu^2a^2$. For
fluctuations about the $\chi$ field, only the zero mode is present, and the
continuum starts at the value $\mu^2a^2$.

For the second pair of solutions, the Schr\"odinger operators corresponding
to fluctuations about the normal modes can be written as
\be
\bar{S}^{(2,\pm)}_2=-\frac{d^2}{dx^2}+5\mu^2a^2\mp4\mu^2a^2\tanh(\mu ax)-
6\mu^2a^2{\rm sech}^2(\mu ax).
\ee
In this case we see \cite{mfe53} that these fluctuations have zero modes and
no other bound states. Furthermore, the continua start at $\mu^2a^2$, and are
formed by reflecting states in the interval $\mu^2a^2$ and $9\mu^2a^2$, and
free states for energies greater than $9\mu^2a^2$.

Before ending this paper, we notice that fluctuations about the first pair of
solutions are described by reflectionless potentials, for which the continua
start at $9\mu^2a^2$ and at $\mu^2a^2$, as shown in Fig.2.

\begin{eqnarray}
\qbezier(-90,20)(-80,-28)(-70,20)
\put(-110,40){\line(-1,0){20}}
\qbezier(-90,20)(-92,40)(-110,40)
\put(-50,40){\line(1,0){20}}
\qbezier(-50,40)(-68,40)(-70,20)
\qbezier(70,20)(80,-28)(90,20)
\put(60,25){\line(-1,0){20}}
\qbezier(70,20)(68,25)(60,25)
\put(100,25){\line(1,0){20}}
\qbezier(100,25)(92,25)(90,20)\nonumber
\end{eqnarray}
\begin{center}
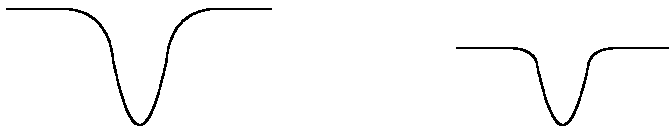
Fig.2. Potentials for the first pair of solutions.
\end{center}

On the other hand, fluctuations about the second pair of solutions are
described by potentials that accommodate reflecting states, and these
reflecting states appear in between the values $\mu^2a^2$ and $9\mu^2a^2$,
as depicted in Fig.3.

\begin{eqnarray}
\qbezier(70,20)(80,-28)(90,20)
\put(60,25){\line(-1,0){20}}
\qbezier(70,20)(68,25)(60,25)
\put(110,40){\line(1,0){20}}
\qbezier(110,40)(92,40)(90,20)
\qbezier(-90,20)(-80,-28)(-70,20)
\put(-60,25){\line(1,0){20}}
\qbezier(-60,25)(-68,25)(-70,20)
\put(-110,40){\line(-1,0){20}}
\qbezier(-90,20)(-92,40)(-110,40)\nonumber
\end{eqnarray}
\begin{center}
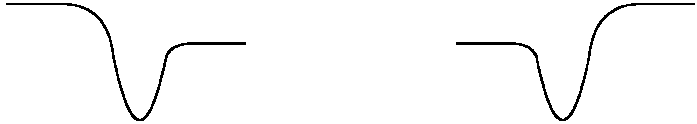
Fig.3. Potentials for the second pair of solutions.
\end{center}

This property in fact independs of the particular ratio between the two
parameters: for $\lambda/\mu >1$, it is not hard to show that fluctuations
about the second pair of solutions are always described by potentials that
accommodate reflecting states in between $\mu^2a^2$ and $4\lambda^2 a^2$,
which are exactly the values where start the continua of the reflectionless
potentials that describe fluctuations about the first pair of solutions.

\section{A General System}
\label{sec:general}

Let us now consider another system, defined by the potential
\be
\label{eq:raj}
U(\phi,\chi)=\frac{1}{4}(\phi^2-1^2)^2+\frac{1}{2}f\chi^2+
\frac{1}{4}\lambda \chi^4+\frac{1}{2}d(\phi^2-1)\chi^2.
\ee
This is the pontential considered in \cite{bca89}, and here we
use the same
notation, with $\lambda$, $f$ and $d$ real and positive.
In this case we see that
\be
U(\phi,0)=\frac{1}{4}(\phi^2-1)^2
\ee
and so this system presents the pair of solutions
\be
\label{eq:pair3}
\bar{\phi}_3(x)=\tanh(x/\sqrt{2}),\qquad
\bar{\chi}_3=0.
\ee
In \cite{bca89} it was shown that this pair is unstable, at least in some
region in parameter space. We recall that the investigation carryed
out in \cite{bca89} was mainly concerned with stability of another pair of
solutions, of the same form of the one given by the second
pair of solutions of the former example. In that investigation, it was also
shown \cite{bca89} that when the parameters
of the system allows for a normal mode diagonalization that leads to
analytical results, the pair of solutions given by the above
Eq.~{(\ref{eq:pair3})} is classically unstable.

Our main interest here is simpler, and concerns investigating if there is
some region in parameter space where
the above pair of solutions is classically or linearly stable. In this case,
the Schr\"odinger operators corresponding to small fluctuations $\eta_n(x)$
and $\xi_n(x)$ described by $\phi(x,t)= \bar{\phi}_3+\eta_n(x) \cos(w_n t)$
and $\chi(x,t)=\bar{\chi}_3+\xi_n(x)\cos(w_n t)$ can be cast to the following
forms
\begin{eqnarray}
\bar{S}_2^{(3,1)}&=&-\frac{d^2}{dx^2}+2-3\,\,{\rm sech}^2(x/\sqrt{2}),
\\
\bar{S}_2^{(3,2)}&=&-\frac{d^2}{dx^2}+f-d\,\,{\rm sech}^2(x/\sqrt{2}).
\end{eqnarray}
The first operator presents two bond states: The zero mode and another bound
state at $w^2=3/2$. However, for the second operator we see that a set of
$n=0,1,2,...$ bound states can appear, where $n$ obeys
\be
n<\frac{1}{2}[\sqrt{1+8d\,}-1].
\ee
This shows that the number of bound states depends only on the parameter
$d$, and that there is at least one bound state for $d>0$. The energy of the
bound states are given by
\be
w^2_n=f-\frac{1}{8}\Bigl[\sqrt{1+8d\,}-1-2n\Bigr]^2\, .
\ee
To avoid instability we focus attention on the deepest bound state: Here we
see that to ensure stability we have to impose the restriction
\be
f\ge\frac{1}{4}\Bigl[1+4d-\sqrt{1+8d}\Bigr],
\ee
and this shows that there is room to choose $d$ and $f$, keeping the
corresponding pair of solutions stable. Furthermore, we remark that since
$1+4d$ is always greater than $\sqrt{1+8d}$ for $d>0$, one can not set
$f\to0$ because this would introduce instability, unavoidably.

On the other hand, stability does not impose any restriction on the sign of
$d-f$, and this allows going a little further on this issue since for $d-f>0$
the above potential presents other minima. To see this explicitly,
let us note that the potential $(\ref{eq:raj})$ also gives
\be
U(0,\chi)=\frac{1}{4}-\frac{1}{2}(d-f)\chi^2+
\frac{1}{4}\lambda \chi^4.
\ee
Here we see that for $d-f\le 0$ the points $(\phi^2_0=1,\chi=0)$ are the only
possible global minima of the potential. However, for $d-f>0$ there are other
minima, at $\phi=0$ and
\be
\chi^2_0=\frac{d-f}{\lambda}
\ee
and these minima can be local or global minima,
depending on the value of the other parameter, $\lambda$.
For simplicity, let us suppose that $\lambda\ge(d-f)^2$. In this case the
points $(\pm1,0)$ are always global minima, but when $\lambda=(d-f)^2$ there
are also global minima at $(0,\pm\sqrt{1/(d-f)})$. 

For $d-f=r>0$, and for $\lambda\ge r^2$ we can get another pair of solutions,
of the same type of the former one, given 
by 
\be
\bar{\phi}_4=0\qquad
\bar{\chi}_4(x)=\sqrt{\frac{r}{\lambda}\,}\,
\tanh\left(\sqrt{r/2\,}
\,\,x\,\right).
\ee
To investigate stability we procced as before: Here the Schr\"odinger
operators are
\begin{eqnarray}
\bar{S}_2^{(4,1)}&=&-\frac{d^2}{dx^2}+
\Bigl[\frac{d}{\lambda}\,r-1\Bigr]-
\frac{d}{\lambda}\,r\,\,{\rm sech}^2\left(\sqrt{r/2\,}\,\,x\,\right)\\
\bar{S}_2^{(4,2)}&=&-\frac{d^2}{dx^2}+2r-3r\,\,
{\rm sech}^2\left(\sqrt{r/2\,}\,\,x\,\right).
\end{eqnarray}
We see that $\bar{S}_2^{(4,2)}$ is like $\bar{S}_2^{(3,1)}$, that is, it
presents the zero mode and another bound state at $w^2=(3/2)r$. However, for
the other Schr\"odinger operator the number of bound states are now
controlled by
\be
n<\frac{1}{2}\Biggl[\sqrt{1+8\frac{d}{\lambda}\,}-1\Biggr],
\ee
and so there is at least one bound state. From the energy of the deepest
bound state we get, to ensure stability
of the corresponding pair of solutions,
\be
\frac{d}{\lambda}-\frac{1}{r}\ge\frac{1}{4}
\Biggl[1+4\frac{d}{\lambda}-\sqrt{1+8\frac{d}{\lambda}\,}
\,\,\Biggr].
\ee
This restriction implies that $\lambda<dr$. Now, if we write
$\lambda=sr^2$ we obtain
\be
1\le s<\frac{d}{r}=\frac{d}{d-f},
\ee
and this implies that there are many possibilities of choosing $s$
without destroying stability of the corresponding pair of solutions. In
particular, we can choose $s=1$ to write $\lambda=r^2=(d-f)^2$, which
shows that in this case the potential $(\ref{eq:raj})$ presents four
degenerate global minima: Two at $\chi=0$ and $\phi^2=1$, and two
at $\phi=0$ and $\chi^2=1/r=1/(d-f)$.

The above results show that when the system presents global minima at the
four points $(\pm1,0)$ and $(0,\pm1/\sqrt{r})$ we can have stable solitons
joining the minima $(\pm1,0)$ by a straight
line with $\chi=0$ or the minima $(0,\pm1/\sqrt{r})$ by another straight line
with $\phi=0$. This is interesting, at least within the context
of searching for defects inside defects, as recently considered in
\cite{mor95}, in the case of systems of the type considered in this section,
and in \cite{brs96a}, in the case of systems belonging to the class of systems
introduced in Sec.~{\ref{sec:systems}}. But this is another issue,
which is presently under consideration.

\acknowledgments
We would like to thank F. Moraes for stimulating discussions, and for
critically reading the manuscript. We also thank F. A. Brito for some
interesting comments.

\end{document}